\documentclass{article}
\usepackage[utf8]{inputenc}
\usepackage[usenames]{color}
\usepackage{colortbl}
\usepackage{graphicx}
\begin{document}
\title{ On viability of inflation in non-minimal kinetic coupling theory}
\author{N. Avdeev$^{1,2}$\thanks{E-mail: NAAvdeev1995@mail.ru} and A.Toporensky$^{2,3}$\thanks{E-mail: atopor@rambler.ru}}
\date{}
\maketitle
$^{1}$Department of Astrophysics and Stellar Astronomy, Faculty of Physics, Lomonosov Moscow State University, Leninskie Gory, 1/2, Moscow 119991, Russia

$^{2}$Sternberg Astronomical Institute, Lomonosov Moscow State University, Universitetsky Prospekt, 13, Moscow 119991, Russia

$^{3}$Kazan Federal University, Kazan 420008, Republic of Tatarstan, Russia
\begin{abstract}
    We consider initial conditions for inflation in non-minimal kinetic coupling theory. If inflation is driven solely by
    kinetic term with no potential, the resulting number of e-folds depends only upon initial velocity of the scalar field
   $ \dot \phi$.
    We write down the expression for the number of e-folds explicitly, and show that for physically reasonable values
    of the coupling constant, we can get 60 e-folds only for exponentially big initial $\dot \phi$. When the scalar field potential is taken into account, the double inflation scenario arises where the first inflation is kinetic term driven, and the second one 
    is potential term driven. In this case
     we need not very large $\dot \phi$ to start with for 60 e-folds, on the other hand,
    some initial condition lead to physically inadmissible eternal inflation. We show numerically that in the  measure used in the present paper
    only smaller part of initial conditions lead to eternal inflation for reasonable values of the coupling constant. 
    
\end{abstract}

Inflation is one of the most popular and successful scenarios explaining the physics of the early universe ~\cite{Starobinsky1,Guth1,Linde1,Albrecht,Linde2,LiLy,Baumann}. The popularity of inflation is mainly due to the fact that it allows us to explain a number of observational data, such as homogeneity, isotropy, and spatial flatness. It also allows us to explain the formation of galaxies with the help of initial quantum oscillations, which are then amplified due to gravitational instability \cite{Mukhanov,Hawking,Starobinsky2,Guth2}.

Despite a success in general, reconstruction of a particular form of inflation which can be in accordance with observational
results on cosmological perturbation is still a problem. Models popular at the beginning of the inflationary paradigm 
(such a massive or self-interactive minimally coupled scalar field) are already ruled out. Current observations of the amplitude
of scalar perturbations as well as upper bounds on the tensor-to-scalar ration indicate that only rather shallow potential is compatible with the minimal coupling. Lifting minimal coupling it is possible to create a viable model for steeper potential,
including the Higgs field. The price for this result is rather high value of the dimensionless constant of non-minimal coupling
\cite{Bezrukov, Bar}
One more viable model, the Starobinsky inflation which uses quadratic corrections to the Einstein gravity and treats the scalar field as an effecting one, also needs big dimensionless coupling constant \cite{Starobinsky1}. In is already known that combination of these two scenarios
into a single model can not reduce values of coupling constants significantly \cite{Star2}.

This motivates further search for other possible scenarios. One of recent proposals deals with a scalar field, non-minimally coupled with gravity by
 its kinetic term. It is interesting that only one type of coupling, the coupling with the Einstein tensor, leads to the second-order equations of motion, so this theory is exceptional within the whole class of kinetic coupling theories
\cite{Sushkov1, Germani1}. Its cosmological dynamics is richer than the dynamics in the case of minimally coupled scalar field. In particular, it allows
inflation even in the absence of the field potential. On the other hand, non-zero potential can induce second  stage
of inflation \cite{I} which either follows the first stage directly or through some transient period \cite{Saridakis}. The goal of the present paper is to find initial conditions leading to an adequate inflation in this scenario. In the current study we use
only one necessary criterion for a successful inflation -- the requirement that number of e-folds is bigger than $60$. This problem allows analytical treatment for zero-potential inflation and requires numerical study in the general case. Unlike earlier works on minimally coupled fields, we do not fix initial energy of the scalar 
field since in the non-minimal coupling case there is some ambiguity in defining scalar field energy
due to presence of cross-terms in the equations of motion.

We work in this units $G=c=\hbar=1$ and use this signature $(-,+,+,+)$.

The action of the theory has the form \cite{Sushkov1}:
\begin{equation}\label{action}
    S = \int d^4 x \sqrt{-g} \bigl( \frac{R}{8\pi} - [g^{\mu\nu}+\kappa G^{\mu\nu}]\phi_{,\mu}\phi_{,\nu}-2V(\phi)\bigr) \\
\end{equation} 
where $R$ is the scalar curvature, $g_{\mu\nu}$ is metric tensor, $G_{\mu\nu}$ is the Einstein tensor, $V(\phi)$ is the scalar potential, $\kappa$ is the coupling parameter.

In the spatially flat  Friedmann-Robertson-Walker cosmological model the action (\ref{action}) yields the following field equations \cite{Sushkov2}.
\begin{equation}\label{equations_1}
    3H^{2} = 4\pi\dot \phi^{2}(1-9 \kappa H^{2}) + 8\pi V(\phi)\\
\end{equation}  

\begin{equation}\label{equations_2}
 2\dot H + 3 H^{2} = -4 \pi \dot \phi^{2} (1 + \kappa (2 \dot H + 3 H^{2} + 4 H\ddot \phi \dot{\phi}^{-1})) + 8\pi V(\phi)\\
\end{equation}

\begin{equation}\label{equations_3}
    (\ddot\phi + 3 H \dot\phi) - 3\kappa(H^{2}\ddot\phi + 2 H \dot H \dot \phi + 3 H^{3} \dot \phi \dot{)} = -V_{\phi} \\
\end{equation}

We start with the simplest case of $V(\phi) = 0$, as in the model (\ref{action}) the kinetic coupling solely can induce
the  inflation. 
The field equations in this case are as follows:
\begin{equation}\label{equations_zero1}
    3H^{2} = 4\pi\dot \phi^{2}(1-9 \kappa H^{2}) \\
\end{equation}

\begin{equation}\label{equations_zero2}
 2\dot H + 3 H^{2} = -4 \pi \dot \phi^{2} (1 + \kappa (2 \dot H + 3 H^{2} + 4 H\ddot \phi \dot{\phi}^{-1})) \\
\end{equation}

\begin{equation}\label{equations_zero3}
    (\ddot\phi + 3 H \dot\phi) - 3\kappa(H^{2}\ddot\phi + 2 H \dot H \dot \phi + 3 H^{3} \dot \phi \dot{)} = 0 \\
\end{equation} 

We can rewrite this system isolating the highest derivative terms

\begin{equation}\label{rewrite_eqzero2}
    \dot H = \frac{-3 H^{2}-4 \pi \dot \phi^{2}(1-9\kappa H^{2})}{2(1+\frac{4 \pi \dot \phi^{2} \kappa (1+9\kappa H^{2})}{1-3\kappa H^{2}})}
\end{equation} 

\begin{equation}\label{rewrite_eqzero3}
    \ddot \phi = -3 H \dot \phi + \frac{3 H \dot \phi \kappa (-3 H^{2}-4\pi \dot \phi^{2}(1-9\kappa H^{2}))}{1-3 \kappa H^{2}+4\pi \dot \phi^{2} \kappa (1+9\kappa H^{2})}
\end{equation} 

\begin{equation}\label{rewrite_eq1}
    3H^{2} = 4\pi\dot \phi^{2}(1-9 \kappa H^{2}) \\
\end{equation} 
which allows us to write down an equation containing $H$ and its derivative only:
\begin{equation}\label{rerewrite_eqzero2}
    \dot H = \frac{-3H^2(1-9\kappa H^2)(1-3\kappa H^2)}{1-9\kappa H^2 + 54\kappa^2 H^4} \\
\end{equation} 
This equation can be solved in a closed form, so we get:
\begin{equation}\label{solve_zero2}
    \sqrt{\kappa}ln((\frac{1+\sqrt{3\kappa}H}{1-\sqrt{3\kappa}H})^{\frac{1}{\sqrt{3}}}(\frac{1-3\sqrt{\kappa}H}{1+3\sqrt{\kappa}H})^{\frac{1}{2}})+\frac{1}{3H} = t + const \\
\end{equation} 
Constant in this equation depends on the initial conditions. Now we introduce  the number of inflationary e-foldings since the onset of inflation (this means that the universe  expands $e^N$ times during inflation)
\begin{equation}\label{N}
    N = log(\frac{a(t_{end})}{a(t_{initial})}) = \int\limits_{t_i}^{t_e} H dt \\
\end{equation}
For realistic inflationary scenarios 
this parameter should satisfy the condition $N_e>60$ at the end of inflation. For the case of the theory with zero scalar potential, one can obtain an analytic expression for $N$. Rewriting   (\ref{rerewrite_eqzero2})
\begin{equation}\label{rerewrite_eqzero2_1}
    dH \frac{1-9\kappa H^2 + 54\kappa^2 H^4}{-3H(1-9\kappa H^2)(1-3\kappa H^2)}= Hdt \\
\end{equation} 
and using (\ref{N}) we obtain
\begin{eqnarray}\label{N_solve}
    N =\int\limits_{t_i}^{t_e} \frac{1-9\kappa H^2 + 54\kappa^2 H^4}{-3H(1-9\kappa H^2)(1-3\kappa H^2)} dH = \nonumber\\ ln(H(t_e)^{-\frac{1}{3}}(1-9\kappa H(t_e)^2)^{\frac{1}{6}}(1-3\kappa H(t_e)^2)^{-\frac{1}{3}}) -\nonumber\\
    ln(H(t_i)^{-\frac{1}{3}}(1-9\kappa H(t_i)^2)^{\frac{1}{6}}(1-3\kappa H(t_i)^2)^{-\frac{1}{3}}) 
\end{eqnarray}

We introduce also the first slow-roll parameter  $\epsilon$ \cite{Linde2}
\begin{equation}\label{eps}
    \epsilon = -\frac{\dot H}{H^2} \\
\end{equation}
which implies that the universe accelerates, $\ddot a > 0$, when $\epsilon_H < 1$.

The indicator of the end of inflation is the condition  $\epsilon_H = 1$. Thus, based on this it is possible to calculate the value of Hubble parameter at the end of inflation. We 
write down the equation for this value using (\ref{eps}) and (\ref{rerewrite_eqzero2})
\begin{equation}\label{eq_for_eps}
    \frac{3(1-9\kappa H(t_{e})^2)(1-3\kappa H(t_{e})^2)}{1-9\kappa H(t_{e})^2 + 54\kappa^2 H(t_{e})^4} = 1, \\
\end{equation}
the solution of this equation is
\begin{equation}\label{finalHubble_1}
    \kappa H(t_{e})^2 = \frac{27-\sqrt{513}}{54}\approx 0.081. \\
\end{equation} 

Now we calculate the initial values for $H$ and $\phi$ at which N would be at least 60 (using eq.(\ref{N_solve})) at the moment when $\epsilon = 1$(see eq.(\ref{finalHubble_1}) and (\ref{eps})). The initial value of the Hubble constant is a solution of the following inequality
\begin{eqnarray}\label{N_solve_ineq}
    ln(H(t_i)^{-\frac{1}{3}}(1-9\kappa H(t_i)^2)^{\frac{1}{6}}(1-3\kappa H(t_i)^2)^{-\frac{1}{3}}) < q - N
\end{eqnarray}
where
\begin{equation}\label{finalHubble_notations2}
    q = ln(H(t_e)^{-\frac{1}{3}}(1-9\kappa H(t_e)^2)^{\frac{1}{6}}(1-3\kappa H(t_e)^2)^{-\frac{1}{3}})
\end{equation}
Here we present some values of $q$ and $H(t_e)$ for different $\kappa$
\begin{eqnarray}\label{q_values}
    \kappa = 1 \    \ q \approx 0.297\    \ H(t_e) \approx 0.284\nonumber\\
    \kappa = 10 \    \ q \approx 0.681\    \ H(t_e) \approx 0.09\nonumber\\
    \kappa = 10^6 \    \ q \approx 2.599\    \ H(t_e) \approx 0.00028\nonumber\\
    \kappa = 10^{12} \    \ q \approx 4.9\    \ H(t_e) \approx 2.84\cdot10^{-7}
\end{eqnarray}
We rewrite this inequality (\ref{N_solve_ineq}) as
\begin{eqnarray}\label{N_solve_ineq1}
    H(t_i)^{\frac{1}{3}}((1-9\kappa H(t_i)^2)^{\frac{1}{6}}-\exp^{q-N}(1-3\kappa H(t_i)^2)^{\frac{1}{3}}H(t_i)^{\frac{1}{3}})\times\nonumber\\
    \times(1-3\kappa H(t_i)^2)^{\frac{1}{3}} < 0
\end{eqnarray}
For obtaining a solution to the inequality, we find the boundary of the admissible interval for H. To do this, we solve this equation
\begin{equation}\label{finalHubble}
    AH(t_i)^6 - BH(t_i)^4 + CH(t_i)^2 - 1 = 0 \\
\end{equation} 
where
\begin{equation}\label{finalHubble_notations1}
    A = 9\kappa^2 e^{6q-6N}; \ \  B = 6\kappa e^{6q-6N}; \ \ C = 9\kappa + e^{6q-6N};\\
\end{equation} 

Thus, we obtain an interval of values for H and $\phi$ at which more than 60 e-folds are accumulated during inflation: $H \in (H_{sol};\frac{1}{3\sqrt{\kappa}})$ and $\phi \in (\phi_{sol}, \infty)$ (see Appendix), note that $H=\frac{1}{3\sqrt{\kappa}}$ is the highest possible value of $H$ in this theory, as it can be easily seen from the Eq. (10).

As we see, $\dot \phi$ must be exponentially large ($\dot\phi\sim 4.97\cdot10^{77}$ for $\kappa = 1$ and $N = 60$; $\dot\phi\sim 4.97\cdot10^{71}$ for $\kappa = 10^{12}$ and $N = 60$) and $H$ must be very close to $\frac{1}{3\sqrt{\kappa}}$ (for $N = 60$ and $\kappa = 1$, $H$  differs from $\frac{1}{3\sqrt{\kappa}}$ by  $1.1\cdot10^{-79}$ and  for $\kappa = 10^{12}$  this difference is $1.1\cdot10^{-91}$). Note that
the value of coupling constant $\kappa$ enters in the expression for $\dot \phi$ only through $q$ which changes very smoothly with
$\kappa$ (see Eq. (\ref{finalHubble_notations2})), so  these big values of $\dot \phi$ can not be easily compensated by changing $\kappa$. So that, only very  specific initial conditions must be met in order to trigger adequate inflation.

The situation changes when we introduce a scalar field potential. Now, same as for the minimally coupled case, the potential itself
can induce inflation. Moreover, the zero potential asymptotic is still a solution if the potential is less steep than the quadratic one.
So that, in the model under investigation we can have two independent inflationary stages, one induced
by the kinetic coupling, and the second stage induced by the potential.
This makes easier to get at least 60 e-foldings during
inflation. However, in this case we are faced with another problem. Namely, it is known that apart from inflation with a natural
exit the system contains another regimes, which are inappropriate for description of our Universe. 
\begin{figure}[ht]
\begin{minipage}[ht]{0.45\linewidth}
\center{\includegraphics[width=1\linewidth]{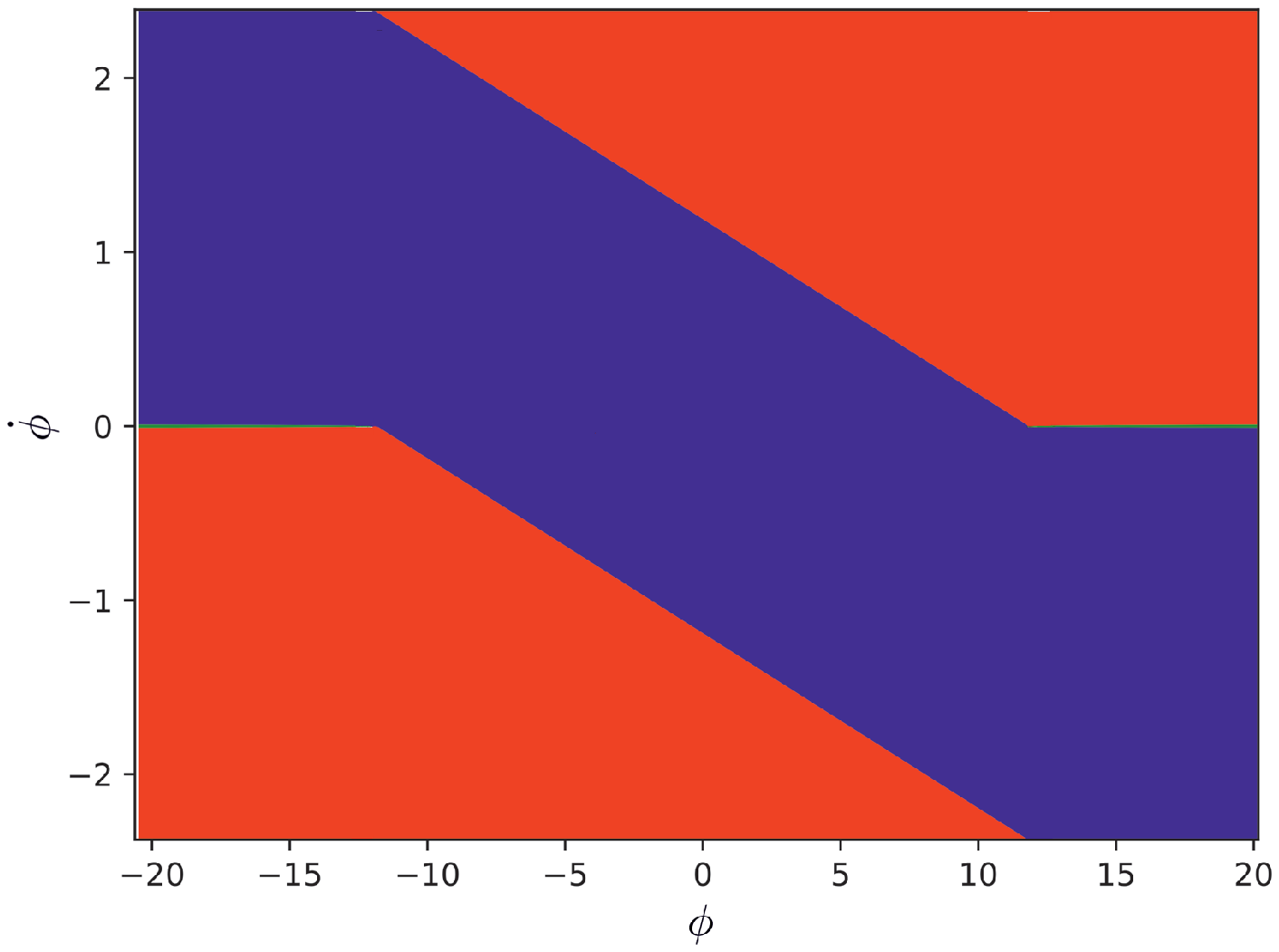} }
\end{minipage}
\hfill
\begin{minipage}[ht]{0.45\linewidth}
\center{\includegraphics[width=1\linewidth]{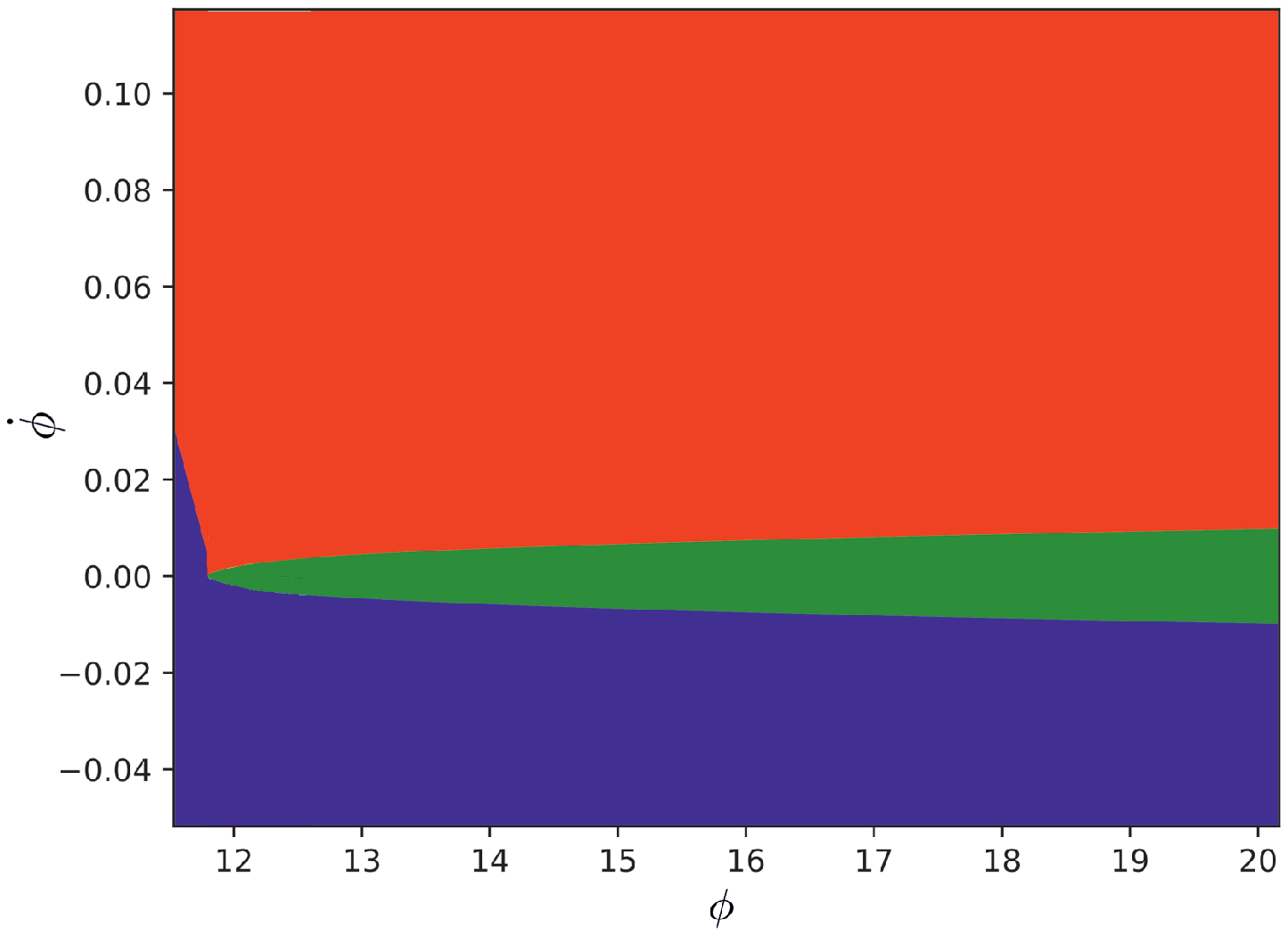} }
\end{minipage}
\hfill
\begin{minipage}[ht]{1\linewidth}
\begin{tabular}{p{0.45\linewidth}p{0.45\linewidth}}
\centering a) & \centering b) \\
\end{tabular}
\end{minipage}
\vspace*{-1cm}
\caption{ The diagrams  are plotted in the  coordinates $\phi$ and $\dot{\phi}$.  The red and green areas show initial conditions for which  the eternal inflation scenario is realized,the blue zone indicates the initial conditions leading to sufficient inflation.  Here $\kappa=100$, $V_0=10^{-5}$.
The right panel is zoomed part of the left plot showing the region with the green area.
}
\label{ris:image0}
\end{figure}
In the present paper we consider a potential which grows less rapidly than the quadratic one, leaving the quadratic (when the first inflation stage can exist but it is modified) 
and steeper potentials (when we have only potential induced inflation) for a future work.
 For definiteness, 
we consider $V = V_0 \phi^{1.5}$.

\begin{figure}[ht]
\center{\includegraphics[scale=0.5]{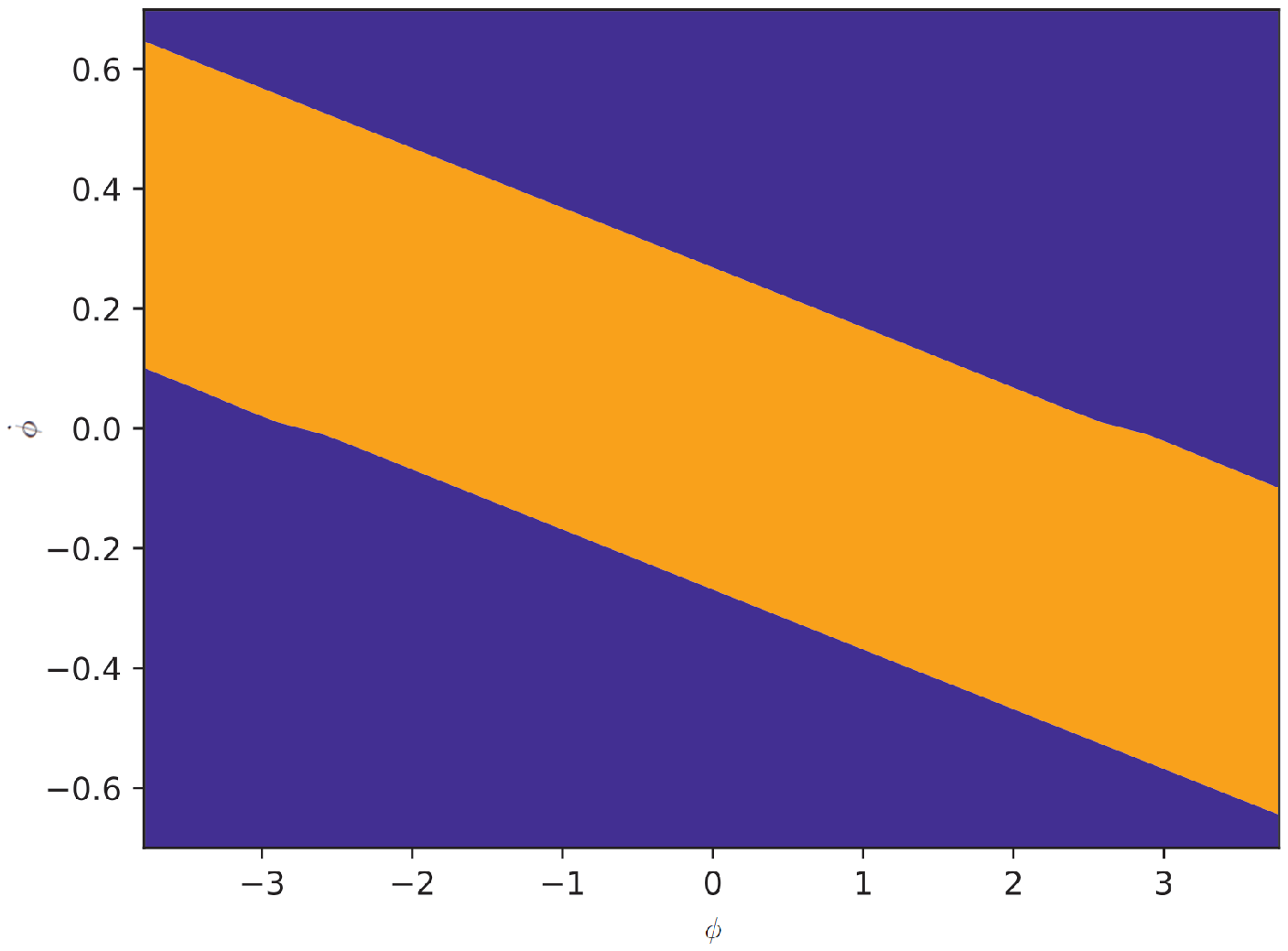}}
\caption{ The blue zone represent initial condition giving sufficient inflation, in the orange zone inflation is absent
or insufficient.}
\label{ris:image2}
\end{figure}

\begin{figure}[ht]
\begin{minipage}[ht]{0.45\linewidth}
\center{\includegraphics[width=1\linewidth]{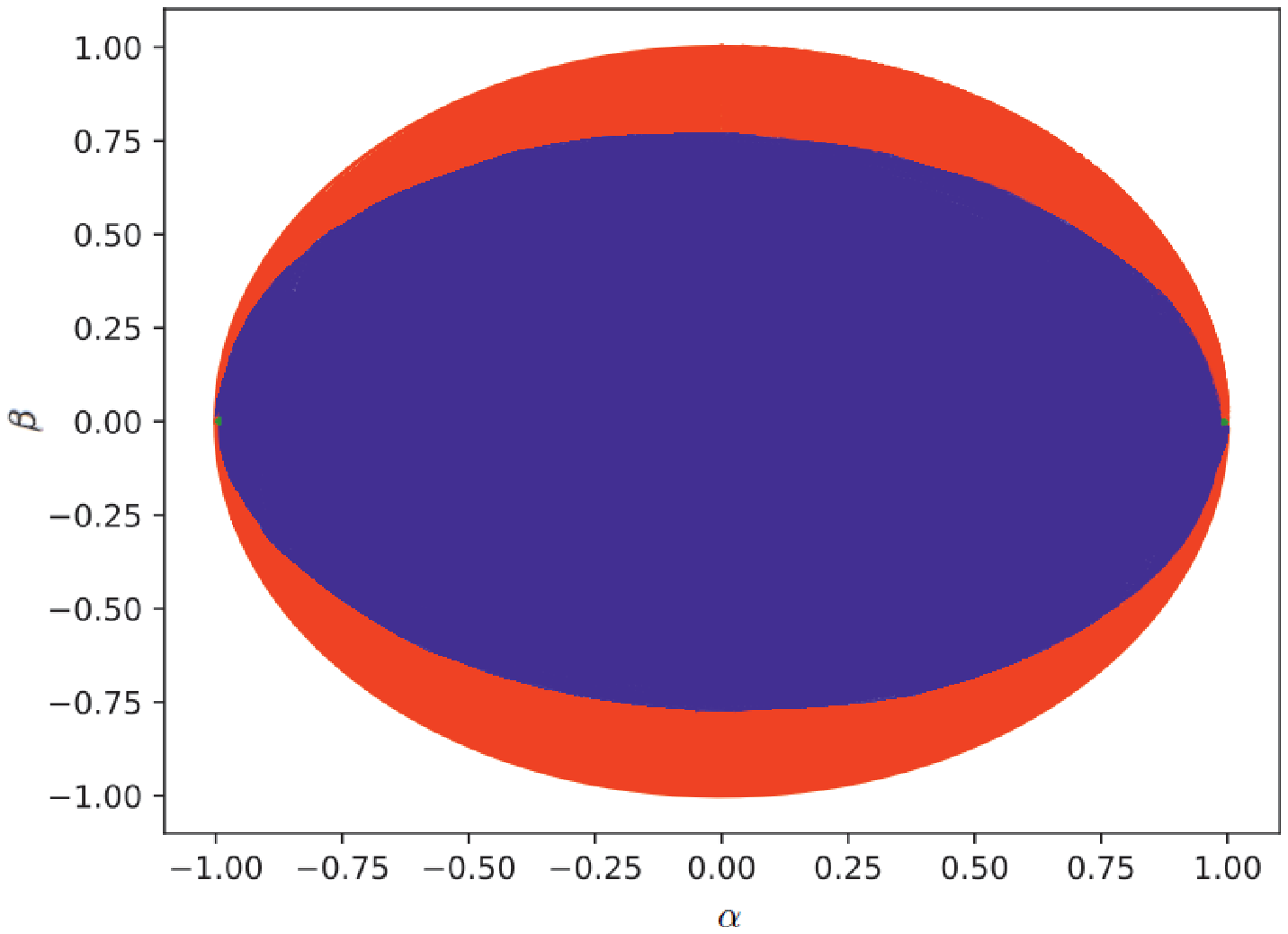} }
\end{minipage}
\hfill
\begin{minipage}[ht]{0.45\linewidth}
\center{\includegraphics[width=1\linewidth]{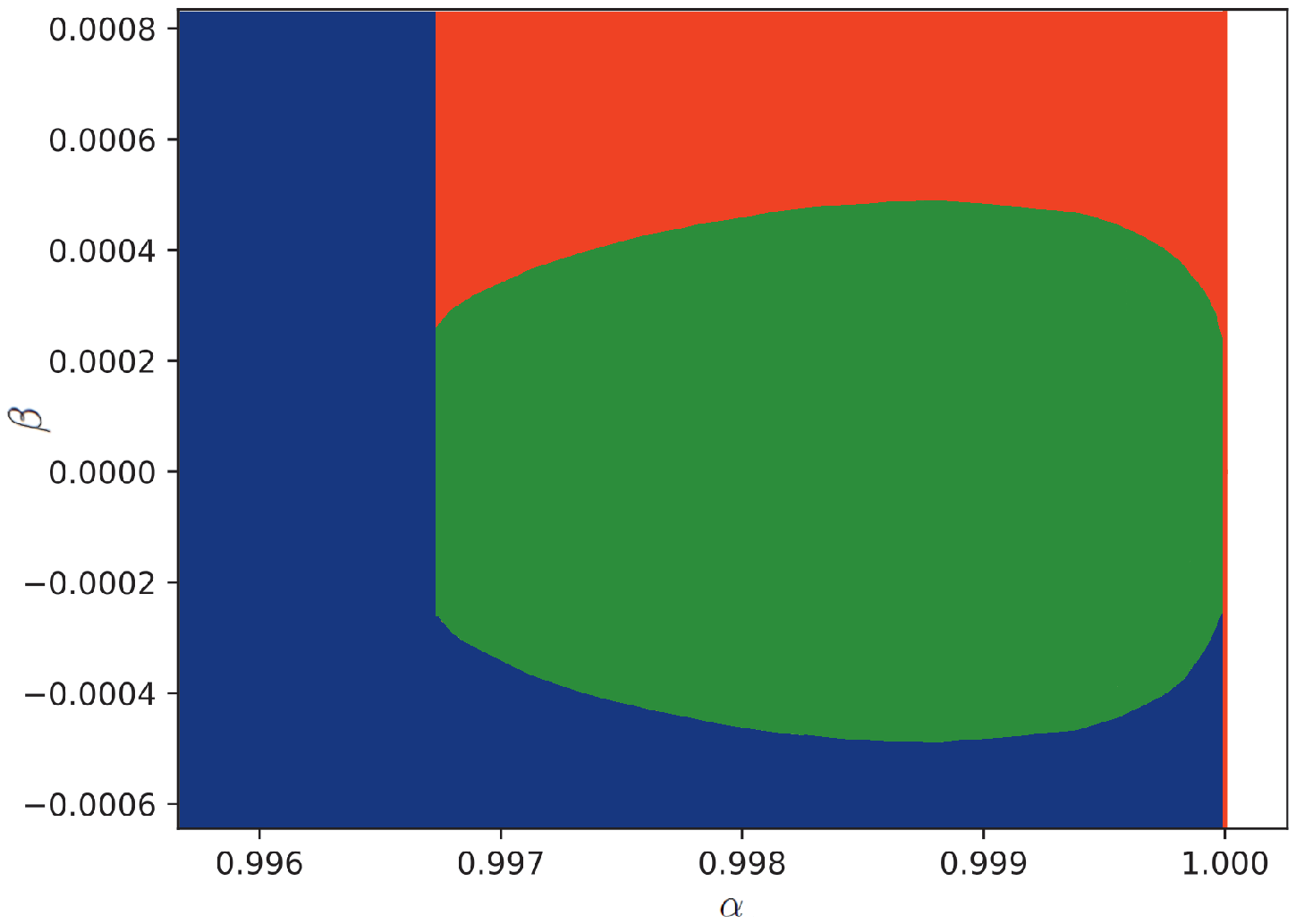} }
\end{minipage}
\hfill
\begin{minipage}[ht]{1\linewidth}
\begin{tabular}{p{0.45\linewidth}p{0.45\linewidth}}
\centering a) & \centering b) \\
\end{tabular}
\end{minipage}
\vspace*{-1cm}
\caption{ The diagrams  are plotted in the  coordinates $\alpha = \frac{\phi}{\sqrt{1 + \phi^2 + \dot \phi^2}}$ and $\beta = \frac{\dot\phi}{\sqrt{1 + \phi^2 + \dot \phi^2}}$.  The red and green areas show initial conditions for which  the eternal inflation scenario is realized,the blue zone indicates the initial conditions leading to sufficient inflation.  Here $\kappa=100$, $V_0=10^{-5}$.
The right panel is zoomed part of the left plot showing the region with the green area.
}
\label{ris:image1}
\end{figure}

Now the  system of equations cannot be solved analytically, however, it is possible to obtain its numerical solutions and find initial conditions leading to adequate inflation. For presentation of numerical results we have chosen $V_0 = 10^{-5}$ and $\kappa = 100$.

The initial conditions space of the model studied is divided into three zones depending on the output of the dynamical evolution.
 In the first zone, the scenario of double inflation and the subsequent exit from  the second inflation is realized.  The blue area on the Fig.1 represents this zone. This zone is divided into two subzones, shown in the Fig.2. The first subzone (orange zone) represents the initial conditions leading to less than 60
e-folds.
 And in the second subzone (blue zone), the number of e-folds exceeds 60, so these initial conditions lead to adequate inflation.

In the second and third zones in the Fig.1 the accelerated expansion never stops.  These two zones correspond to two different asymptotics.  
  In the second zone (marked in green in Fig.1) the scalar field  $\phi$ and Hubble parameter $H$ tend to (see \cite{Toporensky1,Jiro1,Jiro2})
\begin{equation}\label{assimptotic2}
    \phi(t)_{t\rightarrow\infty} \approx \biggr[\frac{\sqrt{V_0}}{4}t+const\biggl]^4;\  \
    H_{t\rightarrow\infty} \approx \frac{1}{\sqrt{3\kappa}}
\end{equation}
The initial data from zone 3 lead to asymptotics described in \cite{Jiro1,Jiro2}:
\begin{equation}\label{assimptotic3}
    \phi(t)_{t\rightarrow\infty} \approx \biggr[\frac{33}{64\pi\kappa}\sqrt{\frac{1}{24\pi V_0}}t+const\biggl]^{4/11};\  \
    H_{t\rightarrow\infty} \approx \sqrt{\frac{8\pi V_0}{3}}\phi^{3/4} \sim t^{3/11}
\end{equation}

The configuration of an orange zone in the Fig.2 (insufficient or no inflation) is qualitatively
the same as for the massive minimally coupled scalar field (see \cite{Swagat}), being 
a narrow strip inside a region of sufficient inflation. However, as we can see from the Fig.1, for large enough
values of the initial scalar field two new zones with eternal accelerated 
expansion appear. These zones correspond to regimes inappropriate for the inflationary
scenario. As in the present paper we consider initial conditions of an arbitrary initial
energy, the phase space is not compact, and we can not attribute a reasonable measure to it.
However, by making the transformation 
$$
\alpha = \frac{\phi}{\sqrt{1 + \phi^2 + \dot \phi^2}},
$$
and
$$
\beta = \frac{\dot\phi}{\sqrt{1 + \phi^2 + \dot \phi^2}}. 
$$
we obtain a compact phase space $(\alpha, \beta)$ and Fig. 3 shows that the bigger part of initial
conditions for the physically admissible part $|\alpha|^2+|\beta|^2 \le 1$ of the plane $(\alpha, \beta)$
does not lead to eternal inflation. As for the eternal inflation, the regime (\ref{assimptotic3}) dominates over
the regime (\ref{assimptotic2}) from the viewpoint of initial conditions since the green zone in the  Figs.1,3 is very small.

Our results can be summarized as follows. Though in principle inflation in the model under investigation can be driven
by kinetic coupling term only, the adequate inflation (no less than 60 e-folds) requires either exponentially large initial
values of $\dot \phi$ or exponentially large values of the coupling constant $\kappa$. Both possibilities are problematic from the 
physical point of view. In particular,  if the theory in question is a low-energy approximation of some more general underlying theory,
the required initial conditions may be located far beyond a zone where this theory can be considered as a reliable approximation.

On the contrary, non-zero scalar field potential naturally leads to a successful inflation without any fine-tuning.
Though in this case the necessary number of e-folds is collected mostly due to scalar field potential (so that, the same way
as in the standard inflationary scenario),
a novel thing
specific for kinetic coupling theories  is that some part of e-folds can be collected
at the stage of kinetic term inflation. In the time interval between these two types of inflation the inflationary parameters
can change rapidly providing significant effects on the propagation  of cosmological perturbations. This effect needs a special
investigation and we leave it for a future research.

\section*{Acknowledgments}
The work of  AT have been supported by the RFBR grant 20-02-00411.
 NA thanks the Foundation for the Theoretical Physics and Mathematics Advancement Foundation “BASIS,” of which he is a fellow. 
 AT thanks  the Russian Government
Program of Competitive Growth of Kazan Federal University.

\section*{Appendix}
We rewrite Eq.(\ref{finalHubble}) in the following notations: $x=H^2$, $a_1=A$, $b_1=-B/3$, $c_1=C/3$ and $d_1=-1$:
\begin{equation}\label{appendix1}
    a_1 x^3 + 3b_1 x^2 + 3c_1 x + d_1 = 0.
\end{equation}
This equation has three solutions that can be found analytically. 
Using the substitutions $y=x + \frac{b_1}{a_1}$, $p_1 = \frac{c_1 a_1 - b^2_1}{a_1^2}$ and $2q_1 = \frac{2b_1^3-3a_1 b_1 c_1 + d_1 a_1^2}{a_1^3}$ we get from the Eq.(\ref{appendix1}) the following equation
\begin{equation}\label{appendix2}
    y^3 + 3p_1 y + 2q_1 = 0.
\end{equation}

Now, using the Cardano formula we get that solutions of equation (\ref{appendix1}) are
\begin{equation}
    x = \bigg(-q_1 \pm \sqrt{q_1^2+p_1^3}\bigg)^{1/3}-\frac{p_1}{\bigg(-q_1 \pm \sqrt{q_1^2+p_1^3}\bigg)^{1/3}}-\frac{b_1}{a_1}.
\end{equation}
In our case, two of them are complex solutions and one is real one.   The real solution 
gives us the following exact expression for the Hubble parameter $H$ through the number of e-folds $N$  and its approximation
for large $N$:

\begin{eqnarray}\label{ap_1}
   H_{sol}^2 = \frac{\bigg(\bigg(36\sqrt{\frac{972k^2-27e^{6(q-N)}\kappa+8e^{12(q-N)}}{e^{6(q-N)}\kappa}}\kappa-324\kappa-8e^{6(q-N)}\bigg)e^{12(q-N)}\bigg)^{2/3}-4e^{6(q-N)}(27\kappa-e^{6(q-N)})}{18e^{6(q-N)}\kappa\bigg(\bigg(36\sqrt{\frac{972\kappa^2-27e^{6(q-N)}\kappa+8e^{12(q-N)}}{e^{6(q-N)}\kappa}}\kappa-324\kappa-8e^{6(q-N)}\bigg)e^{12(q-N)}\bigg)^{1/3}}+\nonumber
\end{eqnarray}
\begin{flushleft}
\begin{equation}
     +\frac{2}{9\kappa}\approx \frac{1}{9\kappa}\bigg(1-\frac{13e^{6(q-N)}}{324\kappa}\bigg)
\end{equation}
\end{flushleft}
Now $\dot\phi_{sol}$ takes the form
\begin{equation}\label{ap_6}
\dot\phi_{sol} = \sqrt{\frac{3H_{sol}^2}{4\pi (1-9\kappa H_{sol}^2)}}\sim \frac{3\sqrt{3}}{\sqrt{13\pi}}\cdot e^{3(N-q)}
\end{equation}

\end{document}